# Static and dynamic critical behaviour of 3d random site Ising model: different Monte Carlo algorithms

**D. Ivaneyko [a], J. Ilnytskyi [b], B. Berche [c,d], and Yu. Holovatch [b,c,a]**

[a] Ivan Franko National University of Lviv, 79005 Lviv, Ukaine
[b] Institute for Condensed Matter Physics, National Academy of Sciences of Ukraine, 79011 Lviv, Ukraine
[c] Laboratoire de Physique des Matériaux, Université Henri Poincaré, Nancy 1, 54506 Vandœuvre les Nancy Cedex, France
[d] Instituto Venezolano de Investigaciones Científicas, Centro de Física, Aptdo 21827, 1020-A Caracas, Venezuela

We perform numerical simulations to study static and dynamic critical behaviour of the 3d random-site Ising model. A distinct feature of our approach is a combination of the Metropolis, Swendsen-Wang, and Wolff Monte Carlo algorithms. For the static critical behaviour, these approaches are the complementary ones, whereas in dynamics they correspond to three different types of relaxation, being a particular subject of our study.

Key words: random Ising model, Monte Carlo, critical exponents

Influence of structural disorder on the critical behaviour is a long-standing and still unsettled problem of condensed matter physics. A particular problem we rise in this report is the critical behaviour of the site diluted quenched 3d Ising model (random Ising model, RIM). Such a model is of relevance for description of criticality of different systems with scalar order parameter. For fluids, it is an archetype model to study liquid-gas critical point of a fluid in porous medium [1]. It describes a cubic lattice of $N$ sites, $N_1$ of them are occupied by one-component vectors, the rest being empty, an interaction is imposed between occupied sites only. For the concentrations of occupied sites $p=N_1/N$ above the percolation threshold, a second-order phase transition occurs in such a system and this transition is characterized by critical exponents differing from those of the regular (undiluted) 3d Ising model (see Tables 1 and 2 for recent data). Numerous analytical, numerical and experimental studies have addressed peculiarities of the RIM criticality [1]. In particular, the Monte Carlo (MC) simulations have been performed using different algorithms. In the present analysis, we made use of three different algorithms, Metropolis, Swendsen-Wang and Wolff ones, which enabled us to find out regions of their applicability and to discriminate between them choosing an optimal one.

Local dynamics MC algorithms (e.g., in Metropolis form) being used near the critical point suffer of the critical slowing down problem. The simple explanation is that the large clusters of correlated spins cause most single-spin flips attempts to be wasted. The cluster algorithms with non-local dynamics (in Swendsen-Wang or Wolff form) have been developed to overcome this difficulty. Effectively, these algorithms utilise mapping of the phase transition phenomena onto a percolation problem of spin connectivity, the critical temperature of the former problem corresponding to the percolation threshold of the latter and, as a consequence, non-local cluster updates are performed. However, being equivalent in terms of equilibrium properties, local and cluster algorithms describe in fact different dynamical systems and the corresponding dynamics of latter algorithms has not been studied theoretically yet. In this paper we treat cluster algorithms not only as an efficient simulational tools but also study their dynamics for RIM, particularly concentrating on the dynamical critical index $z$. This is also interesting from the point of view of the general physical considerations relating $z$ to certain combinations of static critical indices [2].

We performed MC simulations at the critical temperature of infinite diluted spin system on $L \times L \times L$ simple cubic lattice with periodic boundary conditions. The maximum lattice size $L=96$ and concentration of occupied sites $p=0.85$ were chosen to minimize the crossover phenomena. Our MC simulations consisted of two independent parts. The first part of simulations aimed on an estimate of typical autocorrelation times and static critical exponents. Our analysis showed the Wolff algorithm to be the most efficient one, therefore the data for static exponents are obtained from the simulations, based on this algorithm [3]. Typically, we used $10^3$ disorder realizations. For each realization $250\,\tau_E$ equilibration and $10^3\,\tau_E$ production MC steps are used ($\tau_E$ being the autocorrelation time). In the second part, we concentrated on the critical dynamics of all three MC algorithms [4] with approximately the same statistics. Particular attention was paid to an accurate account of the large-time behaviour of autocorrelation functions.

Table 1
RIM static critical exponents for the correlation length, susceptibility, and order parameter. RG: field-theoretical renormalization group calculations, MC: Monte Carlo simulations. Data for the regular 3d Ising model are given in the square brackets.

|        | $\nu$ | $\gamma$ | $\beta$ |
|--------|-------|----------|---------|
| RG     | $0.678(10)^5$ | $1.330(17)^5$ | $0.349(5)^5$ |
|        | $[0.6304(13)^6]$ | $[1.2396(13)^6]$ | $[0.3258(14)^6]$ |
| MC     | $0.683(3)^7$; $0.68(2)^8$ | $1.342(6)^7$; $1.34(1)^8$ | $(0.3535(17)^7$; $0.35(1)^8$ |
|        | $[0.631(1)^9]$ | $[1.2390(71)^{10}]$ | $[0.3269(6)^9]$ |
| MC[3]  | $0.662(2)$ | $1.314(4)$ | $0.337(1)$ |

Table 2
RIM relaxation time critical exponents for local, Swendsen-Wang, and Wolff dynamics. RG: field-theoretical renormalization group calculations for the model A. MC: Monte Carlo simulations. Data for the regular 3d Ising model are given in the square brackets.

|        | $z_{LD}$ | $z_{SW}$ | $z_W$ |
|--------|----------|----------|-------|
| RG     | $2.18^{11}$; $2.237^{12}$ | | |
|        | $[2.017^{13}]$ | | |
| MC     | $2.4(1)^{14}$; $2.62(7)^{15}$ | $0.41^8$ | |
|        | $[2.032(4)^{16}]$ | $[0.59^8$; $0.50^{17}]$ | $[0.28(2)^{18}$; $0.33(1)^2]$ |
| MC[4]  | $2.26(2)$ | $0.36(2)$ | $0.19(1)$ |

The results of our simulations are presented in last lines (MC) of Tables 1 and 2. Whereas the static exponents (Table 1) distinctly differ from those of the pure Ising model, some of them did not reach the asymptotic RIM values. This may be caused by the fact that for the dilution and lattice sizes chosen the system still crosses over to an asymptotic regime. Dynamical exponents (Table 2) characterize three different types of dynamics, as explained above. For the local dynamics our result $z_{LD}$ is in a good agreement with available data; $z_{SW}$ and $z_W$ are calculated for the site-diluted model for the first time. Our value for $z_{SW}$ is in a reasonable agreement with its counterpart for the random-bond Ising model calculated at the bond dilution $p=0.7$ [8].

Let us note again that the exponents $z_{SW}$ and $z_W$ have not been yet calculated theoretically neither for the pure nor for the random models. However, whereas in the pure model they can be related to the combination of static exponents [2], the very existence of such a relation in the diluted case remains unsettled.

We acknowledge useful discussions with C. Chatelain and W. Janke.